\documentstyle[twoside,fleqn,espcrc2]{article}

\let\a=\alpha
\let\b=\beta 
\let\c=\gamma
\let\C=\Gamma  
\let\d=\delta
\let\e=\epsilon
\def\eb{{\bar\epsilon}}
\def\tb{{\bar\theta}}
\def\kb{{\bar\kappa}}
\def\etab{{\bar\eta}}

\def\tv{\tilde v}

\let\k=\kappa
\let\l=\lambda
\let\L=\Lambda
\let\m=\mu
\let\n=\nu 
\let\r=\rho
\let\s=\sigma
\let\S=\Sigma
\let\t=\theta
 
\def\hm{\hat{\mu}}
\def\hn{\hat{\nu}}  
\def\ha{\hat{\alpha}} 
\def\hb{\hat{\beta}}

\def\sh#1{\rlap{\hbox{$\mskip 1 mu /$}}#1}	

\def\p{\partial}
\def\bd{\begin{document}} 
\def\ed{\end{document}}
\def\be{\begin{equation}}
\def\ee{\end{equation}}
\def\ba{\begin{array}}
\def\ea{\end{array}}
\def\bea{\begin{eqnarray}}
\def\eea{\end{eqnarray}}
\def\nn{\nonumber}
\def\ni{\noindent} 
\let\la=\label 
\let\bl=\bigl 
\let\br=\bigr
\let\Br=\Bigr 
\let\Bl=\Bigl
\let\bm=\bibitem
\def\ft#1#2{{\textstyle{{\scriptstyle #1}
\over {\scriptstyle #2}}}}
\def\fft#1#2{{#1 \over #2}}
\def\sst#1{{\scriptscriptstyle #1}}
\newcommand{\eq}[1]{(\ref{#1})}
\def\eqs#1#2{(\ref{#1}-\ref{#2})}
\def\cites#1#2{\cite{#1}-\cite{#2}}
\def\Hat#1{\widehat{#1}}
\def\nosum{({\rm no\ sum\ over}\ i~)}

\def\pl#1#2#3{Phys.~Lett.~{\bf {#1}B} (19{#2}) #3}
\def\np#1#2#3{Nucl.~Phys.~{\bf B{#1}} (19{#2}) #3}
\def\cqg#1#2#3{Class.~and Quant.~Gr.~{\bf {#1}} (19{#2}) #3}
\def\mpl#1#2#3{Mod.~Phys.~Lett.~{\bf A{#1}} (19{#2}) #3}
\def\pr#1#2#3{Phys.~Rev.~{\bf {#1}D} (19{#2}) #3}
\def\prep#1#2#3{Phys.~Rep.~{\bf {#1}C} (19{#2}) #3}

\newcommand{\hoch}[1]{$\, ^{#1}$}

\newcommand{\ttbs}{\char'134}
\newcommand{\AmS}{{\protect\the\textfont2
  A\kern-.1667em\lower.5ex\hbox{M}\kern-.125emS}}

\begin{document}

\title{Supersymmetry in Dimensions Beyond Eleven}


\author{
I. Rudychev, E. Sezgin and P. Sundell 
\address{Center for Theoretical Physics,\\
\hspace{0.08cm} Texas A\& M University, College Station, TX 77843, USA} 
}

\begin{abstract}

Spacetime superalgebras with 64 or less number of real supercharges,
containing the type IIB Poincar\'e superalgebra in $(9,1)$ dimensions
and the $N=1$ Poincar\'e superalgebra in $(10,1)$ are considered. The
restriction $D\le 14$, and two distinct possibilities arise: The
$N=(1,0)$ superalgebra in $(11,3)$ dimensions, and the $N=(2,0)$
superalgebra in $(10,2)$ dimensions. Emphasizing the former, we describe
superparticle and super Yang-Mills systems in $(11,3)$ dimensions. We
also propose an $N=(2,1)$ superstring theory in $(n,n)$ dimensions as a
possible origin of super Yang-Mills in $(8+n,n)$ dimensions. 

\end{abstract}

\maketitle


\section{ Introduction }


It is by now well known that the type IIA string in ten dimensions is
related to M-theory on $S_1$, and the $E_8\times E_8$ heterotic string
is related to M-theory on $S^1/Z_2$ (See \cite{pkt1} for a review).
However, the connection between M-theory and type IIB theory, $SO(32)$
heterotic string and the type I string theory is less direct. One needs
to consider at least a dimensional reduction to nine dimensions to see
a connection. 

One may envisage a unification of the type IIA and type IIB strings in
the framework of a higher than eleven dimensional theory. The simplest
test for such an idea is to show that the Poincar\'e superalgebras of the
IIA/B theories are both contained in a spacetime superalgebra in 
$D\ge 10$ dimensions. The downside of this reasoning is an old result due to
Nahm \cite{nahm}, who showed that, with certain assumptions made,
supergravity theories are impossible in more than $(10,1)$ dimensions
(and supersymmetric Yang-Mills theories in more than $(9,1)$
dimensions)
%
\footnote {A candidate supermultiplet in $(11,1)$ dimensions was
considered in \cite{pvn}, but it was shown that no corresponding
supergravity model exists. }.
%
He assumed Lorentzian signature, and required that no spin higher than
two occurs. Much later, an analyis of super p-brane scan allowing
spacetimes with non-Lorentzian signature, the possibility of a $(2,2)$
brane in $(10,2)$ dimensions was suggested \cite{duff}. 

More recently, various studies in M-theory have also indicated the
possibility of higher than eleven dimensions
\cite{hull,vafa,km1,at,b1,pkt2}. (See also,
\cite{jr,b2,ns,b3,b4,es,n1,b5,rs,b6,b7,n2,b8,n3,rs2}). In most of these
approaches, however, one needs to introduce constant null vectors into
the superalgebra which break the higher dimensional Poincar\'e symmetry.
Accordingly, one does not expect the usual kind of supergravity theory
in higher than eleven dimensions \cite{hull,vafa}. 

An approach which maintains higher dimensional Poincar\'e symmetry has
been proposed \cite{b2}. However, much remains to be done to determine
the physical consequences of this approach \cite{b3,b4}, since it
requires a nonlinear version of the finite dimensional super-Poincar\'e
algebra, in which the anticommutator of two supercharges is proportional
to a product of two or more translation generators \cite{b2} (see also
\cite{es,b5})
\footnote {
While the occurrence of nonlinear terms in a {\it finite}
dimensional algebra is unusual, we refer to \cite{db} for a study of
finite W-algebras where such nonlinearities occur (see also
\cite{w1,w2,w3}). It would be interesting to generalize this study to
finite super W-algebras that contain nonlinear spacetime superalgebras
of the kind mentioned here. 
}.
Simplest realizations of these types of algebras involves 
multi-particles, as was shown first for bosonic systems 
in \cite{b4,b6}, and later for superparticles in \cite{rs,b7,b8,rs2}.
Putting all particles but one on-shell yields an action for a
superparticle in which the constant momenta of the other particles
appear as null vectors.

In what follows, we shall focus our attention on the superalgebraic
structures in $D>11$ that may suggest a IIA/B unification and their
field theoretic realizations which involves null vectors, or certain
tensorial structures, explicitly. We shall then summarize and extend the
results of \cite{es,rs} for the three-superparticle system and its
coupling to $(11,3)$ dimensional super Yang-Mills. Finally, we shall
outline the structure of an $N=(2,1)$ superstring theory in $(n,n)$
dimensions as a possible origin of super Yang-Mills in $(8+n,n)$
dimensions.


\section{Unification of IIA/B Algebras in $D > 10$}


To begin with, let us recall the properties of spinors and Dirac
$\c$-matrices in $(s,t)$ dimensions where $s(t)$ are the number of
space(time) coordinates. The possible reality conditions are listed in
Table 1, where $M, PM, SM, PSM$ stand for Majorana, pseudo Majorana,
symplectic majorana and pseudo symplectic Majorana, respectively
\cite{kt}. An additional chirality condition can be imposed for $s-t=
0~{\rm mod}~4$.
 
The symmetry properties of the charge conjugation matrix
$C$ and the $\c$-matrix $(\c^\m C)_{\a\b}$ are listed in
Table 2. The parameters $\e_0$ and $\e_1$ arise in the relation
\bea
&& C^T=\e_0 C\ ,\nn\\
&& (\c^\m C)^T = \e_1 (\c^\m C)\ . \la{sp}
\eea
This information is sufficient to deduce the 
symmetry of $( \c^{\m_1\cdots \m_r} C)_{\a\b}$ for any $r$, since 
the symmetry property alternates for $r~{\rm mod}~2$.

Using Tables 1 and 2, it is straightforward to deduce the structure of
the type IIA/B superalgberas in $(9,1)$ dimensions. The $N=(1,1)$ super
Poincar\'e algebra (i.e. type IIA) contains a single $32$ component
Majorana-Weyl spinor generator $Q^\a$, with $\a=1,...,32$ and a set of
$528$ bosonic generators, including the translation generator $P^\m$,
that span a symmetric $32\times 32$ dimensional symmetric matrix. The
non-trivial part of the algebra reads
\be
{\bf D=(9,1)\ ,   IIA:}  \la{2a}
\ee
\bea
\{ Q_\a, Q_\b\} &=& \c^\mu_{\a\b}\ P_\mu + (\c_{11})_{\a\b}\ Z
        		+(\c_{11}\c^\mu)_{\a\b}Z_\mu \nn\\
&& +\c^{\mu\nu}_{\a\b}\ Z_{\mu\nu}
       + (\c_{11}\c^{\mu_1...\mu_4})_{\a\b}\ Z_{\mu_1...\mu_4}\nn\\
&& +\c^{\mu_1...\mu_5}_{\a\b}\ Z_{\mu_1...\mu_5}\ . \nn
\eea

All the generators labelled by $Z$ in this algebra, and all the
algebras below, commute with each other. The generators of the Lorentz
group can be added to all these algebras, and the $Z$-generators
transform as tensors under Lorentz group, as indicated by their indices.

It is clear that the algebra \eq{2a} can be written in a $(10,1)$
dimensional covariant form
\be
 {\bf D=(10,1)\ ,   N=1:} \la{11d}
\ee
$$
\{ Q_\a, Q_\b\}=\c^\mu_{\a\b}\ P_\mu +\c^{\mu\nu}_{\a\b}\ 
Z_{\mu\nu}\ +\c^{\mu_1...\mu_5}_{\a\b}\ Z_{\mu_1...\mu_5}\ . 
$$


\begin{table}
\begin{tabular}{|c|c|c|c|}
\hline
 {}  &    {}  &      {}      &       {}       \\
$\e$ & $\eta$ & $(s-t)\,{\rm mod}\,8$ &   ${\rm Spinor Type}$  \\
\hline\hline
 $+$   &    $+$   &   0,1,2      &       M       \\
\hline
 $+$   &    $-$   &    0,6,7      &       PM      \\
\hline
 $-$   &    $+$  &    4,5,6      &       SM      \\
\hline
 $-$   &    $-$  &    2,3,4      &       PSM      \\
\hline
\end{tabular}
\vspace{0.3cm}
\caption{Spinors in (s,t) Dimensions}
\end{table}


\begin{table}
\begin{tabular}{|c|c|c|}
\hline
{} &{}  & {} \\
$t\,{\rm mod}\,4$ & $\e_0$ & $\e_1$ \\
\hline\hline
0   & $+\e$      &   $+\e\eta$      \\
\hline
1   & $-\e\eta$  &   $+\e$        \\
\hline
 2   & $-\e$     &   $-\e\eta$   \\
\hline
 3   & $+\e\eta$  &   $-\e$        \\
\hline
 \end{tabular}
\vspace{0.3cm}
\caption{Symmetries of $C$ and $\c^\m C$, see \eq{sp}. }
\end{table}

 
Next, we consider the $N=(2,0)$ algebra in $(9,1)$ dimensions which
contains two Marona-Weyl spinor generators $Q_\a^i\, (\a=1,...,16, \,
i=1,2)$ and $528$ bosonic generators. The nontrivial part of this algebra
takes the form 
\be
{\bf D=(9,1)\ ,   IIB: } \label{2b}
\ee
\bea
\{ Q_\a^i, Q_\b^j\} &=&\tau^{ij}_a\left(\c^\mu_{\a\b}\ Z_\mu^a +
                \c^{\mu_1...\mu_5}_{\a\b}\ Z_{\mu_1...\mu_5}^{a+}\right) \nn\\
             && +\e^{ij}\c^{\mu\nu\rho}_{\a\b}\  Z_{\mu\nu\rho}\ , \nn
\eea
where $\tau_a$ are the $2\times 2$ matrices $\tau^a= (\s_3,\s_1, 1)$. We
can make the identification $Z_\mu^3\equiv P^\mu$. 

Various aspects of the algebras discussed above have been treated in
\cite{pkt3} in the context of brane charges, and in \cite{b5} in the
context of higher dimensional unification of IIA/B superalgebras. 

In order to unify the superalgebras \eq{11d} and \eq{2b}, we consider
superalgebras in
\be
D=(10+m,1+n)\ , \qquad m,n=0,1,...\la{dims}
\ee
dimensions. To simplify matters, we shall restrict the number real
supercharges to be
\be
{\rm dim}~Q \le 64\ .
\ee
Using Table 1 and Table 2, we learn that this restiction requires
dimensions \eq{dims} with
\be
m+n \le 3\ . 
\ee
Examining all dimensions \eq{dims} for which $(m,n)$ obey this
condition, we find that there are two distinct possibilities: 
\begin{itemize}
\item{$N=(1,0)$ algebra in $D=(11,3)$}    
\item{$N=(2,0)$ algebra in $D=(10,2)$}  
\end{itemize}

Both of these algebras contain $64$ real supercharges, and the second
one is not contained in the first.

The $N=(2,0)$ algebra in $(10,2)$ dimensions has two Majorana-Weyl spinor
generators $Q_\a^i\ (\a=1,...,32,\ i=1,2)$ obeying the anticommutator
\be 
{\bf D=(10,2)\ , N=(2,0): } \la{12}
\ee
\bea
\{Q_\a^i,Q_\b^j\} =& \tau_a^{ij}\left( \c^{\mu\nu}_{\a\b}\, Z_{\mu\nu}^a +
\c^{\mu_1...\mu_6}_{\a\b}\, Z_{\mu_1...\mu_6}^{a+}\right)\nn\\
& +\e^{ij}\left( C_{\a\b}\, Z + \c^{\mu_1...\mu_4}_{\a\b}\, 
Z_{\mu_1...\mu_4}\right) \ .\nn 
\eea 
The $N=(1,0)$ algebra in $D=(11,3)$ dimensions, on the other hand, takes
the form
\be
{\bf D=(11,3)\ , N=(1,0): } \la{14}
\ee
$$
\{Q_\a, Q_\b\} = (\c^{\m\n\r})_{\a\b}~Z_{\m\n\r}
+ (\c^{\m_1\cdots \m_7})_{\a\b}~Z^+_{\m_1\cdots \m_7}\ .
$$ 
Here and in \eq{12}, the $\c$-matrices are chirally projected. Factors
of $C$ used to raise or lower indices of $\c$-matrices are suppressed
for notational simplicity. Note also that both \eq{12} and \eq{14} have
$2080$ generators on their right hand sides, spanning $64\times 64$
dimensional symmetric matrices.

Various dimensional reductions of the algebra \eq{14} yield:
\begin{itemize}
\item[--]{ $N=1$ algebras in $D=(11,2),\ (10,3)$ }
\item[--]{ $N=1$ algebra in $D=(11,1)$ }
\item[--]{ $N=(1,1)$ algebra in $D=(10,2)$ }
\item[--]{ $N=2$ algebra in $D=(10,1)$ }
\end{itemize}
all of which have 64 real supercharges, and contain the $(9,1)$
dimensional IIB and $(10,1)$ dimensional $N=1$ algebras. The algebra \eq{12}
reduces to the last one in the above list. 

The master algebras \eq{12} and \eq{14} also give the $N=1$ algebra in
$D=(9,3)$, the $N=2$ algebra in $D=(9,2)$, the $N=(2,2)$ and $N=(2,1)$
algebras in $D=(9,1)$, all of which contain the IIA/B algebras of
$D=(9,1)$, but not the $N=1$ algebra of $D=(10,1)$.

We conclude this section by showing the embedding of the IIA/B and
heterotic algebras of $D=(9,1)$ in the master algebras \eq{12} and
\eq{14}. 

In the case of \eq{12}, the spinor of $SO(10,2)$ decomposes
under $SO(10,2)\rightarrow SO(9,1)\times SO(2)$ into two left-handed
spinors $Q_{\a}^{i}$ and two right-handed spinors $Q^{\a i}$. We are
using chiral notation in which lower and upper spinor indices refer to
opposite chiralities and there can be no raising or lowering of these
indices. We keep only $Z_{\hm \hn}^{a}$ ($\hm=0,1,...,11$) for
simplicity, and make the ansatz
$Z_{\hm\hn}^{a}\s_{a}^{ij}=P_{[\hm}v_{\hn]}^{ij}$. In the reduction to
$(9,1)$ dimensions we set $P_{\hm}=(P_{\m};0,0)$ and
$v_{\hm}^{ij}=(\vec{0};v_{r}^{ij})$, where $r=10,11$. Now the non-vanishing part of the algebra
\eq{12} reads:
\bea
\{Q_{\a}^{i},Q_{\b}^{j}\}&=&\c^{\m}_{\a\b}P_{\m}v_{+}^{ij} \ ,\\
\{Q^{\a i},Q^{\b j}\}&=&(\c^{\m})^{\a\b}P_{\m}v_{-}^{ij} \ , 
\eea
where $v_{\pm}={1\over 2}(v_{10}\pm v_{11})$. The desired embeddings are
then obtained by setting 
\begin{itemize}
\item[IIB:] { $\quad v_{+}^{ij}=\delta^{ij} \ ,\quad v_{-}^{ij}=0 \ .$}
\item[IIA:] { $\quad v_{+}^{ij}=v_{-}^{ij}={1\over 2}(1+\s^{3})^{ij} \ ,$}
\item[Het:] { $\quad v_{+}^{ij}={1\over 2}(1+\s^{3})^{ij} \ , \quad v_{-}^{ij}=0 \ .$}
\end{itemize}

In the case of \eq{14}, the spinor of $SO(11,3)$ decomposes under
$SO(11,3)\rightarrow SO(9,1)\times SO(2,2)$ into two left-handed spinors
$Q_{\a A}$ and two right-handed spinors $Q^{\a}_{ \dot{A}}$, where
$A,\dot{A}=1,2$ label left- and right-handed spinors of $SO(4)$. We keep
only $Z^{\hm\hn\hat{\r}}$ ($\hm=0,1,...,13$) for simplicity, and make
the ansatz $Z_{\hm\hn\hat{\r}}=P_{[\hm}F_{\hn\hat{\r}]}$
%
\footnote {See \cite{b5} which achives the embeddings of the full IIA/B
algebras \eq{2a} and \eq{2b}, by using multi $F$-tensors and taking into
account $Z^{\hm_1...\hm_7}$ }.
%
We now set $P_{\hm}=(P_{\m};0,0,0,0)$, $F_{\m\n}=0=F_{\m r}$ and
$F_{rs}=
(\s_{rs})^{AB}v_{AB}+(\s_{rs})^{\dot{A}\dot{B}}v_{\dot{A}\dot{B}}$,
where $r,s=10,11,12,13$ and $\s$-matrices are the van der Waerden
symbols of $SO(4)$. Now the non-vanishing part of the algebra \eq{14}
reads:
\bea
\{Q_{\a A},Q_{\b B}\}&=& \c^{\m}_{\a\b}P_{\m}v_{AB} \ ,\\
\{Q_{\a \dot{A}}, Q_{\b \dot{B}}\}&=&
\c^{\m}_{\a\b}P_{\m}v_{\dot{A}\dot{B}} \ .
\eea
The desired embeddings are then obtained by setting 
\begin{itemize}
\item[IIB:] { $\quad \det (v_{AB})\neq 0 \ , 
\quad v_{\dot{A}\dot{B}}=0 \ ,$}
\item[IIA:] { $\quad v_{AB}=u_{(A}u_{B)} \ , 
\quad v_{\dot{A}\dot{B}}=u_{(\dot{A}}u_{\dot{B})} \ ,$}
\item[Het:] { $\quad v_{AB}=u_{(A}u_{B)} \ , 
\quad v_{\dot{A}\dot{B}}=0 \ ,$}
\end{itemize}
where $u_{A}$, $u_{\dot{A}}$ are constant spinors. Note that in the case
of IIB, the symmetric matrix $v_{AB}$ has rank two and it can be chosen
to be $\d_{AB}$ in a suitable basis.

For the heterotic case, which we will focus on in the rest of this
paper, the embedding can be equally well realized by choosing (dropping
the hats)
\be
Z_{\m\n\r}=P_{[\m} v_{\n\r]}\ , \la{z}
\ee
where
\be
v_{\m\n} \equiv n_{[\m} m_{\n]}\ , \la{vmn}
\ee
and $n, m$ are constant and mutually orthogonal null vectors:
\be
m^\m m_\m=0\ , \quad n^\m n_\m=0\ ,\quad  m^\m n_\m=0\ . \la{nm}
\ee
With these choices, it is clear that the matrix $\{Q_\a, Q_\b\}$ in
\eq{14} has rank $16$, as appropriate for the heterotic algebra. 


\section{Superparticle Action and Super Yang-Mills
Equations in $(11,3)$ Dimensions}


The simplest brane action in which the symmetry algebra \eq{14},
\eq{vmn},\eq{nm} may be realized is that of a $0$-brane, i.e. a
superparticle \cite{rs,b7}
%
\footnote {We shall not the treat the realizations of the covariant
algebra \eq{14} here, but we refer the reader to \cite{rs,b8,rs2} for
their multi-superparticle realizations.}.
%
In \cite{rs}, an action for superparticle in the background of a second
and third superparticle was obtained essentially by putting the
background superparticles on-shell. The null vectors $m_\m$ and $n_\m$
satisfying \eq{nm} are the constant momenta of the second and third
superparticles. The following action for a superparticle in $(11,3)$ dimensions
was derived \cite{rs}:
\be
I=\int d\tau \left[ -\ft12 e P^\m P_\m 
+ P_\m \left(\Pi^\m -A n^\m - B m^\m \right)\right]\ , \la{in3}
\ee
where $A$ and $B$ are Lagrange multiplier fields similar to the einbein
$e$, and 
\be
\Pi^\m = \p_\tau X^\m-\ft16\tb \c^{\m\n\r}\p_\tau \t~v_{\n\r}\ .\nn\\
\ee
The action is invariant under the bosonic $\xi, \L$ and
$\Sigma$-transformations 
\bea
&&\d e = \p_\tau \xi\ ,\qquad \d A = \p_\tau \L\ , \qquad 
	\d B = \p_\tau \Sigma\ , \nn\\
&&\d X^\m = \xi P^\m +\L n^\m + \Sigma m^\m \ , \la{gt3}
\eea
and the global supersymmetry transformations
\bea
&&\d_\e X^\m = \ft1{12}~\eb\c^{\m\n\r} \t~v_{\n\r} \ , 
\qquad  \d_\e\t =\e\ , \nn\\
&&  \d_\e P^\m = 0\ , \qquad \d_\e A=0\ , \qquad \d_\e B=0\ . \la{isusy3} 
\eea

The action is also invariant under the {\it local} fermionic $\k$,
$\eta$ and $\omega$ transformations 
\bea
\d\t &=& \c^\m P_{\m}\kappa + \c^\m n_\m \eta
	+ \c^\m m_\m \omega\ , \nn\\
\d X^\m &=& \ft16 \tb\c^{\m\n\r} v_{\n\r} \left(\d_\kappa \t+\d_\eta\t 
	+\d_\omega \t\right)\ , \nn\\
\d P^\m &=& 0\ , \nn\\
\d e &=& -\ft23 (\kb \c^{\m\n} \p_\tau \t) v_{\m\n}\ , \nn \\
\d A &=&  -\ft13 (\kb \c^{\m\n} \p_\tau \t)m_\m P_\n 
	-\ft13 (\etab \c^{\m\n} \p_\tau \t) v_{\m\n}\ , \nn \\
\d B &=&  -\ft13 (\kb \c^{\m\n} \p_\tau \t) P_\m n_\n 
	-\ft13 (\kb \c^{\m\n} \p_\tau \t) v_{\m\n}  \nn\\
     && -\ft13 ({\bar \omega}\c^{\m\n} \p_\tau \t) v_{\m\n} \ . \la{tr3} 
\eea

Superparticle actions have also been constructed in \cite{b7}. Our
results essentially agree with each other. Some apparent differences in
fermionic symmetry transformations are presumably due to field
redefinitions and symmetry transformations proportional to the equations
of motion \cite{rs2}.

We next describe the coupling of Yang-Mills background. To this end, it
is convenient to work in the second-order formalism. Elimination of
$P^\m$ in \eq{in3} gives 
\be
I_0=  \ft12 \int d\tau~e^{-1} \Pi^\m \left( \Pi_\m -A n_\m 
						-B m_\m\right)\ .
\la{ac1}
\ee      
The bosonic and fermionic symmetries of this action can be read off from
\eq{gt3} and \eq{tr3} by making the substitution $P^\m \rightarrow e^{-1}
(\Pi^\m-A n^\m-B m^\m)$. To couple super Yang-Mills background to this
system, we introduce the fermionic variables $\psi^r,\ r=1,...,32$,
assuming that the gauge group is $SO(32)$. The Yang-Mills coupling can
then be introduced as 
\be
I_1= \int d\tau~\psi^r \p_\tau \psi^s \p_\tau Z^M A_M^{rs} ,\la{ac2}
\ee
where $Z^M$ are the coordinates of the $(11,3|32)$ superspace, and $A_M^{rs}$
is a vector superfield in that superspace. 

The torsion super two-form $T^A=dE^A$ can be read from the superalgebra
\eq{12} (with \eq{z} understood):
\be
T^c = e^\a \wedge e^\b\, (\c^{cde})_{\a\b}\, v_{de}\ , \quad\quad  T^\a =0\ ,
\la{tc}
\ee
where the basis one-forms defined as $e^A=d Z^M E_M{}^A$ satisfy
\be
d e^c= e^\a \wedge e^\b\ (\c^{cde})_{\a\b}\ v_{de}\ ,\quad\quad d
e^\a = 0\ , \la{se}
\ee
and $a,b,c,... $ are the (11,3) dimensional tangent space indices.

Using these equations, a fairly standard calculation shows that the
total action $I=I_0+I_1$ is invariant under the fermionic gauge
transformations provided that the Yang-Mills super two-form is given by
\be
F=e^\a \wedge e^b \left[\,  (\c^c\chi)_\a v_{cb} - 2(\c_b\l)_\a\, \right]
+ \ft12 e^a \wedge e^b\
F_{b a}\ , \la{fc}
\ee
and that the transformation rules for $e,A$ and $B$ pick up the extra
contributions. These contributions are determined by the requirement of
the cancellation of the terms proportional to $\Pi^2, \Pi\cdot n$ and
$\Pi\cdot m$, respectively, in the fermionic variation of the total
action. They are easy determine, but as their form is not particularly
illuminating we shall not give them here (see \cite{rs}, for the case of
superparticle in (10,2) dimensions). The fermionic field $\psi^r$ must
be assigned the fermionic transformation
rule
\be
\d  \psi^r =  -\d\t^\a A_\a^{rs} \psi^s\ .
\ee

In \eq{fc}, we have introduced the chiral spinor superfield $\chi_\a$
and the anti-chiral spinor superfield $\l$. These fields and $F$ must
satisfy certain constraints so that the Bianchi identity $DF=0$ is
satisfied. These constraints are \cite{n2}
\bea
&& 
n^a F_{a b} = 0 \ , \ m^a F_{a b} = 0 \ ,\la{s1}\\
&& 
n^a \c_a \l = 0 \ , \  m^a \c_a \l = 0 \ , \la{s2}\\
&&
n^a D_a \l = 0 \ , \  m^a D_a \l = 0 \ , \la{s3}\\
&& 
D_\a\chi^\b= ( \c^{a b} )_\a{}^\b F_{a b} \ ,\la{s4}\\
&&
D_\a \l^\b = \ft14(\c^{abcd})_\a{}^\b F_{ab}\ v_{cd}\ ,   \la{s5}\\
&& 
D_\a F_{a b}  = 2\c^c (D_{[a}\chi)_\a v_{b]c}+4 \c_{[a} (D_{b]}\l)_\a 
\ . \la{s6}
\eea

The above constraints are sufficient to solve the super Bianchi identity
$DF=0$, which can be shown \cite{n2} to yield the the super Yang-Mills
system in (11,3) dimensions \cite{es}. Special $\c$-matrix identities
similar to those required for the existence of the usual super
$p$-branes are not needed here. In showing the vanishing of the term
proportional to $e^\a\wedge e^\b\wedge e^\c$, for example, it is
sufficient to do a Fierz rearrangement, and use the constraints \eq{s2}.
One also find that the spinor superfield $\chi$ is unphysical, as it
drops out the equations of motion. 

The component form of the super Yang-Mills equations are \cite{es}
\bea
&&\c^\m D_\m \l=0\ , \la{nf1}\\
&&D^\s F_{\s[\m} v_{\n\r]} + \ft1{12}\bar\l \c_{\m\n\r} \l = 0 \ .\la{nf2}
\eea
In addition to the manifest Yang-Mills gauge symmetry, these equations
are invariant under the supersymmetry transformations
\bea
\d_\e A_\m &=& \bar\e\c_\m\l \ , \la{nt1} \\
\d_\e  \l &=& -\ft14 \c^{\m\n\r\s} \e F_{\m\n} v_{\r\s} \ , \la{nt2}
\eea
and the extra bosonic local gauge transformation
\be
\d_\Omega A_\m = -v_{\m\n}~\Omega^\n \ ,\quad\quad 
\d_\Omega \l = 0\ , \la{not}
\ee
provided that the constraints \eq{s1}-\eq{s3} hold, and that
\bea
 && v^{\m\n} D_\m \Omega_\n =0\ , \nn\\
 && v_\m{}^\r v_\n{}^\s D_\r \Omega_\s = 0  \ .  \la{nc2}
\eea

The commutator of two supersymmetry transformations closes on shell, and
yields a generalized translation, the usual Yang-Mills gauge
transformation and an extra gauge transformation with parameters
$\xi^\mu$, $\Lambda$, $\Omega^\m$, respectively, as follows:
\be
[ \d_{\e_1}, \d_{\e_2} ] = \d_\xi +\d_\Lambda + \d_\Omega\ ,
\la{nclosure}
\ee
where the composite parameters are given by
\bea
\xi^\m &=&  \bar\e_2 \c^{\m\n\r}\e_1\ v_{\n\r}\ , \la{nxi}\\
\Lambda &=& -\xi^\m\ A_\m\ , \la{nl}\\
\Omega^\m &=& \ft12 \bar\e_2 \c^{\m\n\r}\e_1\ F_{\n\r}\ .  \la{no}
\eea
The global part of the algebra \eq{nclosure} indeed agrees with \eq{14},
\eq{z},\eq{vmn},\eq{nm}. Note the symmetry between the parameters
$\xi^\m$ and $\Omega^\m$. The former involves a contraction with
$v_{\m\n}$, and the latter one with $F_{\m\n}$.

In \cite{es}, an obstacle was encountered in extending the above
construction of super Yang-Mills system to higher than $14$ dimensions.
For example, in $(12,4)$ dimensions, while everthing goes through in
much the same fashion as in $(11,3)$ dimensions, the supersymmetric
variation of the Dirac equation gave rise to a term proportional to
${\bar\l}\l$, which appeared to be nonvanishing, and hence problematic
in obtaining the correct Yang-Mills equation. However, as has been observed 
in \cite{n2}, this term actually vanishes due to the constraints
\eq{s2}. In fact, super Yang-Mills systems in $(8+n,n)$ dimensions, for
any $n\ge 1$ have been constructed in \cite{n2}. More recently, an
action for $(10,2)$ dimensional super Yang-Mills, which can presumably
be generalized to higher dimensions, has also been found \cite{n3}. 


\section{$N=(2,1)$ Superstring in $(n,n)$ Dimensions}


In the previous section we have described super Yang-Mills theory in
higher than $(10,2)$ dimensions. The $(10,2)$ dimensional super
Yang-Mills theory can be derived from a critical heterotic string theory
based on the $N=(2,1)$ superconformal algebra \cite{ov,km1}. In this
section we shall describe the underlying critical string theories of the
super Yang-Mills theories in higher than $(10,2)$ dimensions using a
generalization of the heterotic $N=(2,1)$ string of
\cite{km1,ov,km2,km3}. The model is based on an $N=1$ superconformal
algebra for left-movers in $(8+n,n)$ dimensions and an $N=2$
superconformal algebra for right-movers in $(n,n)$ dimensions. Both
these algebras are extended with null currents
%
\footnote{For a construction which uses null-extended $N=(1,1)$
superconformal algebras realized in $(10,2)$ dimensions, see \cite{ff}.
}.
%
The null-extended $N=1$ algebra is realized in terms of free scalars
$X^{\hm}$ and fermions $\psi^{\hm}$ and makes use of $n-1$
mutually orthogonal null vectors 
\be
v_{\hm}^i~v^{\hm\,j}=0\ , \quad i,j=1,...,n-1, \ \ \hm =1,...,8+2n \ .
\ee
The left-moving $N=1$ algebra is realized as
\bea
T&=& -{1\over 2}~\eta_{\hm\hn}~\p X^{\hm} \p X^{\hn} + {1\over 2}~
\eta_{\hm\hn}~\psi^{\hm} \p \psi^{\hn}\ , \\
G &=& \sqrt {2}~\psi^{\hm}\p X^{\hn} \eta_{\hm\hn}\ ,\\
J^i &=& v_{\hm}^i~\p X^{\hm}\ ,\\
\C^i &=& v_{\hm}^i~\psi^{\hm}\ .
\eea 
The basic OPE's are $X^{\hm}(z) X^{\hn}(0)=-\eta^{\hm\hn}\log z$ and
$\psi^{\hm}(z) \psi^{\hn}(0)=-\eta^{\hm\hn}z^{-1}$. The OPE's of the
energy momentum tensor $T$ has central charge $c=12+3n$, and its OPE's
with the currents $G, J^i, \C^i$ imply that they have conformal spin
${3\over 2},1,{1\over 2}$, respectively. Thus, the ghost anomaly is
$c_g=-26+11-(n-1)\times( 2+1)=-(12+3n)$. 

The null-extended $N=2$ algebra is realized in terms of scalars $X^\m$,
and fermions $\psi^\m$ and makes use of a real
structure $I_\m{}^\n$ in $(n,n)$ dimensions obeying
\be
I_{\m\n}=-I_{\n\m}\ ,\qquad I_\m{}^\rho I_\rho{}^\n=\delta_\m^\n\ ,\quad
\m=1,...,2n , \la{rs} 
\ee
where $I_{\m\n}=I_\m{}^\rho \eta_{\rho\n}$. The real structure has $n$
eigenvectors of eigenvalue $+1$, and $n$ eigenvectors of eigenvalue
$-1$. The crucial property of these eigenvectors that allows us to write 
down a critical algebra in $(n,n)$ dimensions is that the inner
products of two eigenvectors of the same eigenvalue vansihes. Hence, in
particular all the eigenvectors are null. Pick $(n-2)$ of these null
vectors, $\tv_{\m}^r$, say of eigenvalue $+1$. They satisfy 
\bea
I_{\m}{}^\n \tv_\n^r &=&\tv_{\m}^{r}\ , \la{pv}\\
\tv_\m^r~\tv^\m_s &=& 0\ ,   \qquad r,s=1,...,n-2\ . \la{vv}
\eea
The right-moving $N=2$ algebra is then realized as
\bea
\bar{T}&=& -{1\over 2}~\eta_{\m\n}~\bar{\p} 
X^\m \bar{\p} X^\n + {1\over 2}~ \eta_{\m\n}~\psi^\m \bar{\p} \psi^\n\ , 
\\
\bar{G}_\pm &=& {1\over \sqrt{2}}~(\eta_{\m\n}\pm 
I_{\m\n})~\psi^\m\bar{\p} X^\n \ ,
\\
\bar{J} &=& -{1\over 2}~I_{\m\n}~\psi^\m\psi^\n\ ,
\\
\bar{J}^r &=& \tv_\m^r~\bar{\p} X^\m\ ,
\\
\bar{\C}^r &=& \tv_\m^r~\psi^\m\ .
\eea
The energy momentum tensor $\bar{T}$ has central charge $\bar{c}=3n$,
and its OPE's with the currents $\bar{G}_\pm, \bar{J}, \bar{J}^r, \C$
imply that they have conformal spin ${3\over 2},1,1,{1\over 2}$,
respectively. Note that the closure of the alegbra requires the eigen
property of the vectors $\tv^r$ as well as their nullness. Hence, in
this case the ghost anomaly is assumes the critical value
$\bar{c}_g=-26+2\times 11-2-(n-2)(2+1)=-3n$. 

Following the usual BRST quantization scheme one constructs the
right-moving supercharges \cite{km1,km2}:
\be
Q_{\ha}=\oint dz~\S_{gh}~S_{\ha}\ ,
\la{q}
\ee
where $S_{\ha}$ are the right-moving spin fields of $\psi^{\hm}$, and
$\S_{gh}=\exp (-\phi/2-\phi_{1}/2-\cdots-\phi_{n-1}/2)$ is the spin
field of the commuting ghosts. The index $\ha$ labels the Majorana
spinor of $O(8+n,n)$. The single-valuedness of OPE algebra of fermionic
vertex operators in the Ramond sector require that $S_{\ha}$, and
therefore $Q_{\ha}$, are Majorana-Weyl.

BRST invariance requires the null-conditions
\be
\sh{v_{i}}Q =0\ ,  \quad i=1,...,n-1. \la{null}
\ee
The standard form of the target space superalgebra is obtained by
considering the anti-commutator of \eq{q} with its picture-change $Q'$: 
\be
Q'=Z Z_{1}\cdots Z_{n-1} Q\ ,
\la{pic}
\ee
where $Z X=\{Q_{BRST},\xi X\}$ is the picture changing operation built
from the BRST charge and the zero-modes $\xi,\xi_1,...,\xi_{n-1}$ of the
$(\xi,\eta)$ systems used for bosonizing the commuting ghosts. The
supercharges \eq{q} and \eq{pic} obey the algebra 
$\{Q_{\ha},Q'_{\hb}\}=(\sh{v_{1}}\cdots\sh{v_{n-1}}\sh{p})_{\ha\hb}$,
which reduces to
\be
\{Q_{\ha},Q'_{\hb}\}=(\c^{\hm_1...\hm_n})_{\ha\hb}~ 
v^{1}_{\hm_1}\cdots v^{n-1}_{\hm_{n-1}}P_{\hm_n}
\ee
in the BRST-invariant sector. The case of $n=2$ has also been discussed in
\cite{km3}.

The spectrum of states depend on the choices for the null vectors $v_i$
and $\tv^r$. The choices for $\tv^r$ break $SO(n,n)$ down to $SO(2,2)$,
and the choices for $v_i$ break $SO(2,2)$ down to $SO(2,1)$ or less.
Generically, one obtains massless states which assemble into a super
Yang-Mills multiplet in $(8+n,n)$ dimensions, which effectively has the
$8+8$ degrees of freedom of the usual $(9,1)$ dimensional super
Yang-Mills, after all the physical states conditions are imposed (see
\cite{km1,km2} for the $n=2$ case). There is a subtlety in the present
case, however, having to do with the spectral flows induced by the
null-currents in the left-moving sector. They shift the (nonchiral)
$(n,n)$ dimensional momentum with multiples of $\tv_r$. The physical
state conditions then force $\tv_r$ to be orthogonal to $v_i$. 

We do not yet know the exact feature of the target space field theory.
We expect, however, that it will be of the kind studied in \cite{hull2},
where the Yang-Mills field strength satisfies a generalized self-duality
condition. 


\section{Comments}


We started out by considering an algebraic unification of the $(9,1)$
dimensional IIA/B superalgebras in higher dimensions, with emphasis on
$(11,3)$ dimensional $N=(1,0)$ algebra \eq{14}. Having made the choices
\eq{z} and \eq{vmn}, however, we have restricted ourselves to the
embedding of a supersymmetric theory with only $16$ supercharges. While
this is useful in understanding how the null vectors arise in a field
theoric realization, ideally one should seek a master field theoretic
realization in which both IIA and IIB (and hence heterotic) symmetries
are realized according to the suitable choices to be made for the
three-form charge occuring in \eq{14}. 

We have focused our attention on zero-brane and the super Yang-
Mills system it couples to, but the considerations reviewed here apply to
higher branes as well \cite{n1,b6,b7}. 

We have also restricted our attention to the IIA/B unification in a
maximal dimensional spacetime (with $64$ real supercharges), namely
$D=(11,3)$. Null reduction of our results for
multi-superparticles and super Yang-Mills, yield corresponding results
for $N=(1,0)$ supersymmetric models in $(10,2)$ dimensions. However,
$N=(2,0)$ supersymmetric results in $(10,2)$ dimensions cannot be
obtained in this way. In fact, a IIA/B unification in the framework of
the $N=(2,0)$ algebra in $(10,2)$ dimensions does not seem to have
attracted attention previously, and it may be interesting to investigate
this case further. 

One of the dividends of a higher than eleven dimensional unification of
IIA/B systems should be a more manifest realization of various duality
symmetries among the ten dimensional strings/branes. As it has been
stressed in \cite{b1,b2}, these symmetries are to be interpreted as the
similarity transformations of the $64\times 64$ symmetric matrix
$\{Q_\a, Q_\b\}$ which leaves the BPS condition ${\rm det}~\{Q_\a,
Q_\b\} =0$ invariant. An explicit realization of these symmetries at the
level of brane actions would be desirable.

The introduction of structures, e.g. null vectors which break the higher
dimensional Poincar\'e symmetry may give the impression that not much is
gained by a higher dimensional formulation, and that it may amount to a
rewriting of the original theory. This is not quite so, even if one
considers the embedding of a single type of algebra in higher
dimensions, when one considers the null vectors as the averages of
certaing quantities, e.g. momenta, attributed to other branes
co-existing with the brane under consideration, as has been illustrated
in \cite{b3,b4}. Furthermore, as mentioned briefly in the introduction,
there exists now a simple realization of $D>11$ superalgebras which
involve the momenta multi-superparticles \cite{b8,rs2}. These do not
involve constant null vectors and maintain manifest covariance in
$D>11$. The multi-brane extension of these results and the nature of
target space field theories they imply, are interesting open problems.

Another aspect of the theories considered here is that their reductions
to lower than ten dimensions give rise to new kinds of super Yang-Mills
theories which, together with supergravity sector which can be included,
are candidates to be the low energy limits of certain $N=(2,1)$ strings.
The utility of such strings lies in the fact that they provide a unified
picture of various branes, e.g. string and membranes \cite{km1,km2},
resulting from different choices for the null vectors. Here we have
generalized the construction of \cite{km1} to higher than $(2,2)$
dimensional targets, indeed to $(n,n)$ dimensional ones, with $n\ge 2$.
The description of their effective target space models is an interesting
problem. We expect that the field equations in these models are related
to the generalized self-dual Yang-Mills systems studied in \cite{hull2}. 
   
In all the algebras considered here the $Z$-type generators commute with
each other. However, there exist interesting extensions of the
Poincar\'e superalgebra in $(10,1)$ dimensions that includes super
two-form \cite{ma1,ma2}, and super five- form \cite{ma3} generators. The
most general such algebra with $N=1$ supersymmetry in $(10,1)$
dimensions has been called the $M$-algebra. In this algebra, there are
non-vanishing (anti) commutators of super two-form generators. The role
of the charges, some of which are bosonic and some fermionic, has not
been understood yet in the context of $M$-theory. However, we expect
them to play an important role. It would be interesting, therefore, to
determine if the $M$-algebra generalizes to the $(10,2)$ and $(11,3)$
dimensional algebras reviwed here, and if such an algebraic structure
can help in arriving at a more unifying picture of a wealth of
$M$-theory phenomena.

\bigskip


\noindent{\sc Acknowledgements}


\medskip

We thank I. Bars, M.J. Duff and C. Vafa for useful discussions at
various stages of this work. One of the authors (E.S.) would like to
thank the International Center for Theoretical Physics in Trieste for
hospitality. This work was supported in part by the National Science
Foundation, under grant PHY-9722090.

\vfill\eject

\baselineskip=14pt


\end{document}